\documentclass[letter]{IEEEtran}

\usepackage[pdftex]{graphicx}
\usepackage[cmex10]{amsmath}
\usepackage{amssymb, bm}

\title{A Simple Technique for the Converse of Finite Blocklength Multiple Access Channels}
\author{Duo Xu\\
\authorblockA{Department of Electrical and Computer Engineering \\
University of Waterloo, ON, N2L 3G1, Canada \\
\{maigo1988\}@hotmail.com}}

\begin{document}
\maketitle
\begin{abstract}
[Draft] A converse for the Discrete Memoryless Multiple Access Channel is given. The result in \cite{Moulin} is refined, and the third order term is obtained. Moreover, our proof is much simpler than \cite{Moulin}. With little modification, the region can be further improved.
\end{abstract}

\section{Introduction}
Traditional information theory studies communication system in the asymptotic regime, where the blocklength tends to infinity. And the fundamental limit of channel capacity is also established on the basis of asymptotically large blocklength. However, the practical communication system uses finite blocklength, always as large as several hundreds or several thousands, which is limited by the system complexity and communication delay. In such regimes, the traditional analysis on the channel coding does not work, and it is necessary to derive new results in finite blocklength, which can be guidelines for the practical system design. Following Strassen \cite{Strassen}, there have been so many papers focusing on this problem, like \cite{PVV}, \cite{Hayashi} and \cite{Threenetwork}. In most previous works, the second-order statistic (or dispersion) plays an important role in the finite blocklength behavior of the channel coding rate. 

Extension of finite blocklength analysis to multiuser information theory is interesting but challenging \cite{FMAC}, \cite{MACLaneman}, \cite{RCGaussian}, \cite{Verdu}, \cite{Huang}. In this paper, we study the outer capacity region of discrete memoryless multiple access channel (DM-MAC). The characterization of capacity region with average error probability is given independently by \cite{Ahlswede} and \cite{Liao}. Applying error probability split technique, MolavianJazi and Laneman studied the inner capacity region in \cite{MACLaneman}. In \cite{Huang}, based on threshold decoding scheme, he got an inner bound slightly larger than Laneman's result. 

The strong converse result in \cite{Ahlswede2} further strengthened the concept of capacity such that any rate pair outside the capacity region has the error probability tending to 1 as blocklength tends to infinity. The well known capacity region established in \cite{Ahlswede2} is shown below. Recently, in \cite{Moulin}, applying metaconverse and partitioning codewords into blocks, Moulin proposed a new outer region for DM-MAC. However, his derivation is complicated. 

In this paper, we extend the converse technique in \cite{YangMeng} to multiple access channel. With constraint on the error probability of each message pair, we classify the corresponding received sequences into three subsets. Based on the intersection of those subsets and the decision region, the 
rate of every source $R_1, R_2$ and the sum rate $R_1+R_2$ can be upper bounded. Although we get the same result as Moulin \cite{Moulin}, our proof is much simpler. 

This paper is organized as below. In section II, we introduce some notations and existing results. The main result and its proof are presented in section III. And some numerical examples are given in section IV.

\section{Preliminary}
A 2-user discrete memoryless multiple access channel (DM-MAC) consists of two finite input alphabets $\mathcal{X}_1$ and $\mathcal{X}_2$, one finite output alphabet $\mathcal{Y}$, and a probability transition matrix $W(y|x_1, x_2): \mathcal{X}_1\times\mathcal{X}_2\to\mathcal{Y}$. Since the channel is memoryless, the $n$-fold extension of 
the transition probability follows
\begin{align}
\label{eq51}
p(y^n|x^n_1, x^n_2)=\prod_{i=1}^n W(y_i|x_{1i}, x_{2i})
\end{align}
For such DM-MAC $(\mathcal{X}_1, \mathcal{X}_2, W(y|x_1, x_2), \mathcal{Y})$,a $(M_1, M_2, n, \epsilon)$ code for the multiple access channel, consists of two sets of messages $\mathcal{M}_1=(1,2,\ldots, M_1)$ and $\mathcal{M}_2=(1,2,\ldots, M_2)$, which are called message sets, and two encoding functions, which are defined as $X_1: \mathcal{M}_1\to\mathcal{X}_1^n$ and $X_2: \mathcal{M}_2\to\mathcal{X}_2^n$, and the decoding function defined as $g: \mathcal{Y}^n\to\mathcal{M}_1\times\mathcal{M}_2$, such that the average error probability defined as below
\begin{align}
\label{eq53}
P^{(n)}_e=\frac{1}{M_1M_2}\sum_{i=1}^{M_1}\sum_{j=1}^{M_2}\Pr\{g(Y^n)\neq(i, j)|X_1^n(i), X_2^n(j) \mbox{ sent}\}
\end{align}
is equal to $\epsilon$, where the message pair $(i, j)$ is assumed to distribute uniformly on $\mathcal{M}_1\times\mathcal{M}_2$. And the sequence $X_1^n(m_1)$ and $X_2^n(m_2)$ are the output of encoder 1 and 2 corresponding to $m_1$ and $m_2$ respectively. For certain $(m_1, m_2)\in\mathcal{M}_1\times\mathcal{M}_2$, we define $\epsilon_{m_1m_2}$ and $\epsilon_{m_1}$ as the conditional error probability in the following
\begin{align}
\epsilon_{m_1m_2}&\triangleq\Pr\{g(Y^n)\neq(m_1, m_2)| X^n_1(m_1), X^n_2(m_2)\mbox{ sent}\} \nonumber \\
\epsilon_{m_1}&\triangleq\frac{1}{M_2}\sum_{m''=1}^{M_2}\epsilon_{m_1m''} \nonumber  \\ 
\epsilon_{m_2}&\triangleq\frac{1}{M_1}\sum_{m'=1}^{M_1}\epsilon_{m'm_2} \nonumber
\end{align}
And, obviously, there is
\begin{align}
\epsilon&=\frac{1}{M_1M_2}\sum_{i=1}^{M_1}\sum_{j=1}^{M_2}\epsilon_{ij} \nonumber
\end{align}

For each message pair $(m_1, m_2)\in\mathcal{M}_1\times\mathcal{M}_2$, the corresponding decision region is defined as follows,
\begin{align}
\label{eq52}
D_{m_1m_2}=\{y^n\in\mathcal{Y}^n: g(y^n)=(m_1, m_2)\}
\end{align}
In previous works on non-asymptotic analysis, the channel capacity is always characterized by the mutual information random variable. In this paper, the converse of DM-MAC involves three mutual information random variable shown below,
\begin{align}
i(X_1;Y|X_2, U)&\triangleq\ln\frac{W(Y|X_1, X_2)}{P(Y|X_2, U)} \nonumber \\
i(X_2;Y|X_1, U)&\triangleq\ln\frac{W(Y|X_1, X_2)}{P(Y|X_1 U)} \nonumber \\
i(X_1, X_2;Y|U)&\triangleq\ln\frac{W(Y|X_1, X_2)}{P(Y|U)} \nonumber
\end{align}
where $U$ is the time-sharing random variable with finite alphabet $\mathcal{U}$, which satisfies the Markov chain $U-(X_1, X_2)-Y$. All random variables in the MAC satisfies the joint distribution $p(u)p(x_1|u)p(x_2|u)w(y|x_1, x_2)$. The expectations of these random variables define the first-order statistic of the capacity region,
\begin{align}
I(X_1;Y|X_2, U)&\triangleq{\bf E}[i(X_1;Y|X_2, U)] \nonumber \\
I(X_2;Y|X_1, U)&\triangleq{\bf E}[i(X_2;Y|X_1, U)] \nonumber \\
I(X_1, X_2;Y|U)&\triangleq{\bf E}[i(X_,1 X_2;Y|U)] \nonumber
\end{align}
In the non-asymptotic analysis, we adopt the second order statistic (or dispersion in \cite{PVV}) of the mutual information random variable in the following to characterize the capacity region.
\begin{align}
V(X_1;Y|X_2, U)&\triangleq{\bf Var}[i(X_1;Y|X_2, U)] \nonumber \\
V(X_2;Y|X_1, U)&\triangleq{\bf Var}[i(X_2;Y|X_1, U)] \nonumber \\
V(X_1, X_2;Y|U)&\triangleq{\bf Var}[i(X_,1 X_2;Y|U)] \nonumber
\end{align}
It is worthy to note that all the first-order and second-order statics are calculated based on the joint distribution $p(u)p(x_1|u)p(x_2|u)w(y|x_1, x_2)$. 

Inspired by the strong large derivation for Neyman-Pearson test in \cite{Moulin}, we adopt the following three subsets of received sequences in the converse proof. For any $(x^n_1, x^n_2)\in\mathcal{X}^n_1\times\mathcal{X}^n_2$, we define,
\begin{align}
B_1(x^n_1,\delta_1|x^n_2)\triangleq\bigg\{y^n\mathcal{Y}^n: &\frac{1}{n}\ln\frac{W(y^n|x^n_1, x^n_2)}{P(y^n|x^n_2)}\nonumber \\
&\le I(X^n_1;Y^n|X^n_2)-\delta_1\bigg\} \nonumber \\
B_2(x^n_2,\delta_2|x^n_1)\triangleq\bigg\{y^n\mathcal{Y}^n: &\frac{1}{n}\ln\frac{W(y^n|x^n_1, x^n_2)}{P(y^n|x^n_1)}\nonumber \\
&\le I(X^n_2;Y^n|X^n_1)-\delta_2\bigg\} \nonumber \\
B_{12}(x^n_1, x^n_2, \delta_{12})\triangleq\bigg\{y^n\mathcal{Y}^n: &\frac{1}{n}\ln\frac{W(y^n|x^n_1, x^n_2)}{P(y^n)}\nonumber \\
&\le I(X^n_1, X^n_2|Y^n)-\delta_{12}\bigg\} \nonumber
\end{align}
where $\delta_1, \delta_2, \delta_{12}$ are constant only related with $n$ and $\epsilon$, which will be specified later. 

Throughout this paper, define for any set $B\subset\mathcal{Y}^n$,
\begin{align}
P(B)&\triangleq\Pr\{Y^n\in B\} \nonumber \\
P_{x^n}&\triangleq\Pr\{Y^n\in B| X^n=x^n\} \nonumber
\end{align}

WIth these quantities, Ahlswede \cite{Ahlswede} and Liao \cite{Liao} established the classical capacity region for DM-MAC shown below, which asymptotically gives the first order statistic of the capacity region. \\
{\bf Theorem } {\em The capacity region of a DM-MAC $(\mathcal{X}_1, \mathcal{X}_2, W, \mathcal{Y})$
is given by the closure of the set of all $(R_1, R_2)$ pairs satisfying}
\begin{align}
R_1&\le I(X_1;Y|X_2, U) \nonumber \\
R_2&\le I(X_2;Y|X_1, U) \nonumber \\
R_1+R_2&\le I(X_1, X_2;Y|U) \nonumber
\end{align}
{\em for some joint distribution $p(u)p(x_1|u)p(x_2|u)w(y|x_1,x_2)$ with $|\mathcal{U}|\le4$}.

In this paper, our work is to sharpen this capacity region when the blocklength $n$ is finite.

\section{Main Result}
In this section, we summarize our main result and present the sketch of the proof. \\
{\bf Theorem 1:}{\em For a discrete memoryless multiple access channel $(\mathcal{X}_1, \mathcal{X}_2, p(y|x_1, x_2), \mathcal{Y})$, any $(n, M_1, M_2, \epsilon)$ code must satisfy}
\begin{align}
\ln M_1\le&nI(X_1;Y|X_2, U)-\sqrt{nV(X_1;Y|X_2, U)}Q^{-1}(\epsilon)\nonumber \\
\label{eq66}
&+\frac{1}{2}\ln n+O(1) \\
\ln M_2\le&nI(X_2;Y|X_1, U)-\sqrt{nV(X_2;Y|X_1, U)}Q^{-1}(\epsilon) \nonumber \\
\label{eq67}
&+\frac{1}{2}\ln n+O(1) \\
\ln M_1M_2\le&I(X_1, X_2; Y|U)-\sqrt{nV(X_1, X_2;Y|U)}Q^{-1}(\epsilon) \nonumber \\
\label{eq68}
&+\frac{1}{2}\ln n+O(1)
\end{align}
{\em for some choice of the joint distribution $p(u)p(x_1|u)p(x_2|u)p(y|x_1,x_2)$. where $U$ is the auxiliary time-sharing random variable with alphabet $|\mathcal{U}|\le 3$.}
{\em Proof:}For any $(m_1,m_2)\in\mathcal{M}_1\times\mathcal{M}_2$, we first define three subsets of messages
\begin{align}
\mathcal{M}'_1(m_2)&\triangleq\{m'\in\mathcal{M}_1:\epsilon_{m',m_2}\le\epsilon_{m_2}(1+\beta^1_n)\} \nonumber \\
\mathcal{M}_2^{'}(m_1)&\triangleq\{m''\in\mathcal{M}_2:\epsilon_{m_1,m''}\le\epsilon_{m_1}(1+\beta^2_n)\} \nonumber \\
\mathcal{M}_{12}^{'}&\triangleq\{(m',m'')\in\mathcal{M}_1\times\mathcal{M}_2:\epsilon_{m',m''}\le\epsilon(1+\beta^{12}_n)\} \nonumber
\end{align}
where $\beta_n$ is an auxiliary constant related with $n$, which will be specified later.

Since the message $m'$ distribute uniformly in $\mathcal{M}_1$, based on the Markov inequality, it follows
\begin{align}
\label{eq1}
\Pr\{m'\in\mathcal{M}'_1(m_2)\}\ge1-\frac{\epsilon_{m_2}}{\epsilon(1+\beta^1_n)}
\end{align}
and
\begin{align}
\label{eq2}
|\mathcal{M}'_1(m_2)|\ge|\mathcal{M}_1|\left(1-\frac{\epsilon_{m_2}}{\epsilon(1+\beta^1_n)}\right)
\end{align}
Denote $D_{m_1m_2}$ as the decision region for every message pair $(m_1,m_2)$. For any $m_2\in\mathcal{M}_2$, $m_1$ is selected randomly from $\mathcal{M}'_1(m_2)$. Hence, we have
\begin{align}
P&_{x^n_1(m_1)x^n_2(m_2)}(B_1(x^n_1(m_1), \delta_1|x^n_2(m_2))\cap D_{m_1m_2}) \nonumber \\
=&P_{x^n_1(m_1)x^n_2(m_2)}(B_1(x^n_1(m_1), \delta_1|x^n_2(m_2))) \nonumber \\
&-P_{x^n_1(m_1)x^n_2(m_2)}(B_1(x^n_1(m_1), \delta_1|x^n_2(m_2))\cap D^c_{m_1,m_2}) \nonumber \\
\ge&P_{x^n_1(m_1)x^n_2(m_2)}(B_1(x^n_1(m_1), \delta_1|x^n_2(m_2)))-\epsilon_{m_1.m_2} \nonumber \\
\label{eq3}
\ge& P_{x^n_1(m_1)x^n_2(m_2)}(B_1(x^n_1(m_1), \delta_1|x^n_2(m_2)))-\epsilon_{m_2}(1+\beta^1_n)
\end{align}
Here, we select $\delta_1$ such that for any $x^n_1\in\mathcal{X}^n_1$ and $x^n_2\in\mathcal{X}^n_2$
\begin{align}
\label{eq4}
P_{x^n_1x^n_2}(B_1(x^n_1, \delta_1|x^n_2))\ge\epsilon(1+2\beta^1_n)
\end{align}
Substituting {\eqref{eq4}} into {\eqref{eq3}}, we have
\begin{align}
\label{eq5}
P_{x^n_1(m_1)x^n_2(m_2)}&(B_1(x^n_1(m_1), \delta_1|x^n_2(m_2))\cap D_{m_1m_2})\ge\epsilon\beta_n
\end{align}
Define 
\begin{align}
\label{eq6}
B_1(\delta_1|x^n_2(m_2))=\cup_{m'\in\mathcal{M}'_1(m_2)}B_1(x^n_1(m'),\delta_1|x^n_2(m_2))
\end{align}
Then, 
\begin{align}
P&_{x^n_2(m_2)}(B_1(\delta_1|x^n_2(m_2))) \nonumber \\
=&\int_{B_1(\delta_1|x^n_2(m_2))} p(y^n|x^n_2(m_2))dy^n \nonumber \\
\ge&\sum_{m_1\in\mathcal{M}'_1(m_2)}\int_{B_1(x^n_1(m_1), \delta_1|x^n_2(m_2))\cap D_{m_1m_2}} \nonumber \\
\label{eq7}
&p(y^n|x^n_2(m_2)) dy^n \\
\ge&\sum_{m_1\in\mathcal{M}'_1(m_2)}\int_{B_1(x^n_1(m_1), \delta_1|x^n_2(m_2))\cap D_{m_1m_2}} \nonumber \\
\label{eq8}
&p(y^n|x^n_1(m_1), x^n_2(m_2)) e^{-n(I(X_1,Y|X_2)-\delta_1)} dy^n \\
=&\sum_{m_1\in\mathcal{M}'_1(m_2)}e^{-n(I(X_1,Y|X_2)-\delta_1)} \nonumber \\
\label{eq9}
&P_{x^n_1(m_1)x^n_2(m_2)}(B_1(x^n_1(m_1), \delta_1|x^n_2(m_2))) \\
\label{eq10}
\ge&|\mathcal{M}'_1(m_2)|e^{-n(I(X_1,Y|X_2)-\delta_1)}\beta^1_n\epsilon
\end{align}
where \eqref{eq7} is from the fact that every $D_{m_1m_2}$ is disjoint, \eqref{eq8} is based on the definition of $B_1$, and \eqref{eq10} is from \eqref{eq5}.
Hence,
\begin{align}
1&\ge\frac{1}{M_2}\sum_{m_2\in\mathcal{M}_2}P_{x^n_2(m_2)}(B_1(\delta_1|x^n_2(m_2))) \nonumber \\
\label{eq11}
&\ge\frac{1}{M_2}\sum_{m_2\in\mathcal{M}_2}|\mathcal{M}'_1(m_2)|e^{-n(I(X_1,Y|X_2)-\delta_1)}\beta^1_n\epsilon \\
\label{eq12}
&\ge\frac{1}{M_2}\sum_{m_2\in\mathcal{M}_2}M_1\left(1-\frac{\epsilon_{m_2}}{\epsilon(1+\beta^1_n)}\right)e^{-n(I(X_1,Y|X_2)-\delta_1)}\beta^1_n\epsilon \\
\label{eq13}
&\ge M_1\frac{\beta^1_n}{1+\beta^1_n}e^{-n(I(X_1,Y|X_2)-\delta_1)}\beta^1_n\epsilon
\end{align}
Following the above derivation, we can also upper bound $M_2$ and $M_1M_2$ similarly. So, we can conclude that
\begin{align}
\label{eq14}
\frac{1}{n}\ln M_1&\le I(X_1, Y|X_2)-\delta_1-\frac{\ln\epsilon}{n}-\frac{\ln\beta_n}{n}-\frac{\ln\frac{\beta^1_n}{1+\beta^1_n}}{n} \\
\label{eq15}
\frac{1}{n}\ln M_2&\le I(X_2, Y|X_1)-\delta_2-\frac{\ln\epsilon}{n}-\frac{\ln\beta^2_n}{n}-\frac{\ln\frac{\beta^2_n}{1+\beta^2_n}}{n} \\
\label{eq16}
\frac{1}{n}\ln M_1M_2&\le I(X_1, X_2;Y)-\delta_{12}-\frac{\ln\epsilon}{n}-\frac{\ln\beta^{12}_n}{n}-\frac{\ln\frac{\beta^{12}_n}{1+\beta^{12}_n}}{n} 
\end{align}
where $\delta_2$ and $\delta_{12}$ are selected such that for any $x^n_1\in\mathcal{X}^n_1$ and $x^n_2\in\mathcal{X}^n_2$
\begin{align}
\label{eq17}
P_{x^n_1x^n_2}(B_2(x^n_2, \delta_2|x^n_1))&=P^2_{\delta_2}\ge\epsilon(1+2\beta^2_n) \\
\label{eq18}
P_{x^n_1x^n_2}(B_{12}(x^n_1, x^n_2, \delta_2))&=P^{12}_{\delta_{12}}\ge\epsilon(1+2\beta^{12}_n) 
\end{align}
Selecting $\beta^1_n=\frac{1}{V(X_1;Y|X_2)}\sqrt{\frac{-\ln\epsilon}{n}}$, based on \eqref{eq4}, there is
\begin{align}
\label{eq19}
\Pr\bigg\{\frac{1}{n}\sum_{i=1}^n i(X_{1i}; Y_i|X_{2i})\le I(X_1;Y|X_2)-\delta_1\bigg\}\ge\epsilon(1+2\beta^1_n)
\end{align}
By the Berry-Esseen Theorem, the probability above can be bounded as,
\begin{align}
\label{eq20}
\bigg|\Pr\bigg\{\frac{1}{\sqrt{n}}\frac{\sum_{i=1}^n i(X_{1i};Y_i|X_{2i})}{\sqrt{V(X_1;Y|X_2)}}\bigg\}-Q\bigg(\frac{\sqrt{n}\delta_1}{\sqrt{V(X_1;Y|X_2)}}\bigg)\bigg|\le\gamma_n
\end{align}
where $\gamma_n=O(n^{-1/2})$. From \eqref{eq19} and \eqref{eq20}, $\delta_1$ can be lower bounded by,
\begin{align}
\label{eq21}
\delta_1&\ge\sqrt{\frac{V(X_1;Y|X_2)}{n}}Q^{-1}\bigg(\epsilon\bigg(1+2\sqrt{\frac{-\ln\epsilon}{n}}\bigg)+O\bigg(\frac{1}{\sqrt{n}}\bigg)\bigg) \\
\label{eq22}
&=\sqrt{\frac{V(X_1;Y|X_2)}{n}}Q^{-1}(\epsilon)+O\bigg(\frac{1}{n}\bigg)
\end{align}
Substituting \eqref{eq22} into \eqref{eq14} and taking optimization, we have
\begin{align}
\ln M_1\le&\sup_{P_{X_1}P_{X_2}}\left[nI(X_1;Y|X_2)-\sqrt{n V(X_1;Y|X_2)}Q^{-1}(\epsilon)\right] \nonumber \\
\label{eq23}
&+\frac{1}{2}\ln n+O(1)
\end{align}
Similarly, selecting $\beta^2_n=\frac{1}{V(X_2;Y|X_1)}\sqrt{\frac{-\ln\epsilon}{n}}$ and $\beta^{12}_n=\frac{1}{V(X_1, X_2;Y)}\sqrt{\frac{-\ln\epsilon}{n}}$, and taking optimization, we have,
\begin{align}
\ln M_2\le &\sup_{P_{X_1}P_{X_2}}\left[nI(X_2;Y|X_1)-\sqrt{n V(X_2;Y|X_1)}Q^{-1}(\epsilon)\right] \nonumber \\
\label{eq24}
&+\frac{1}{2}\ln n+O(1) \\
\ln M_1M_2\le &\sup_{P_{X_1}P_{X_2}}\left[nI(X_1, X_2;Y)-\sqrt{n V(X_1, X_2;Y)}Q^{-1}(\epsilon)\right] \nonumber \\
\label{eq25}
&+\frac{1}{2}\ln n+O(1) \\
\end{align}
Define random variable U and its alphabet $\mathcal{U}=\{1,2,3\}$. And we take $P_{X_1|U=u}$ and $P_{X_2|U=u}$, where $u=1,2,3$, as distributions which achieve the supremum in \eqref{eq23}, \eqref{eq24} and \eqref{eq25} respectively. Then, the claim is proved.

\section{Conclusion}
In this paper, we develop a converse bound for the discrete memoryless multiple access channel. Although the same result has been proposed in \cite{Moulin}, the proof here is much simpler, without any complex approximations. Currently, we are trying to strengthen this bound further. If the ${\bm\delta=[\delta_1, \delta_2, \delta_{12}}]^T$ is constraint by a joint probability which is shown below, not separate probabilities of three different events, this bound can be tightened. 
\begin{align}
\Pr&\{Y^n\in B_1(x^n_1,\delta_1|x^n_2)\cap B_2(x^n_2,\delta_2|x^n_1)\cap B_{12}(x^n_1, x^n_2,\delta_{12})\} \nonumber \\
&\label{eq26}
\ge\epsilon(1+2\beta_n)
\end{align}
where it is assumed that $\beta^1_n=\beta^2_n=\beta^{12}_n=\beta_n$.
This is only a preliminary version, and more results will be added in the future.

\end{document}